\documentclass[%
 amsmath,amssymb,
preprint,%
]{revtex4-1}

\usepackage{graphicx}
\usepackage{color}
\usepackage{dcolumn}
\usepackage{bm}

\usepackage[utf8]{inputenc}
\usepackage[T1]{fontenc}
\usepackage{mathptmx}
\usepackage{color}

\begin{document}



\title[]{Equilibrium shapes and floatability of static and vertically vibrated heavy liquid drops on the surface of a lighter fluid}

\author{Andrey Pototsky}
 \affiliation{Department of Mathematics, Swinburne University of Technology, Hawthorn, Victoria, 3122, Australia}
\email{apototskyy@swin.edu.au}
\author{Alexander Oron}
\affiliation{Department of Mechanical Engineering, Technion-Israel Institute of Technology, Haifa 3200003, Israel}
\author{Michael Bestehorn}%
\affiliation{Institute of Physics, Brandenburg University of Technology, 03013 Cottbus-Senftenberg, Germany }

\begin{abstract}
A small drop of a heavier fluid may float on the surface of a lighter fluid supported by surface tension forces. In equilibrium, the drop assumes a radially symmetric shape with a circular triple-phase contact line. We show theoretically and experimentally that such a floating liquid drop with a sufficiently small volume has two distinct stable equilibrium shapes: one with a larger and one with a smaller radius of the triple-phase contact line. Next, we experimentally study the floatability of a less viscous water drop on the surface of a more viscous and less dense oil, subjected to a low frequency (Hz-order) vertical vibration. We find that in a certain range of amplitudes, vibration helps heavy liquid drops to stay afloat. The physical mechanism of the increased floatability is explained by the horizontal elongation of the drop driven by subharmonic Faraday waves. The average length of the triple-phase contact line increases as the drop elongates  that leads to a larger average lifting force produced by the surface tension.
\end{abstract}

\maketitle

\section{Introduction}
\label{intro}

 Floatation of a small liquid droplet on the surface of a less dense carrier fluid is a striking manifestation of the strength of tensile forces that exist between immiscible liquid and gaseous phases. When in equilibrium, the drop assumes a radially symmetric shape with a circular triple-phase contact line that separates the drop into a sessile upper cap and a pendant lower part. The quintessential question of determining the largest possible drop volume capable of staying afloat was first formulated and studied over 40 year ago \citep{Hartland1976}. From the theoretical point of view, a stationary shape of a liquid drop can be found by solving the minimal surface problem for liquid menisci, subject to a zero net force condition acting on the triple-phase contact line \citep{Princen1963,Princen1965}. Due to its nonlinear nature and complexity of the boundary conditions, this problem can only be approached numerically for a particular combination of fluids.  Prior to the modern era of personal computers, the computation of stationary drop shapes was a highly tedious task that involved manual manipulation of the tabulated solutions of the minimal surface equations. 
This method was used by \citep{Hartland1976} to determine the maximum possible drop volume as a function of the fluid densities and interfacial tensions. Extending the original work \citep{Hartland1976}, subsequent studies focused on experimental verification of the floatability of water drops on oil surfaces \citep{Phan2012,Phan2014}, addressed the role of the line tension \citep{George2016,Bratukhin2001} and developed a simplified model by assuming that the upper sessile cap of the drop is approximately flat \citep{Bratukhin1994}. 

Here we use numerical continuation method and conduct a series of experiments to reveal a previously unreported multistability of sufficiently small liquid drops floating on the surface of a less dense bulk fluid. Depending on the deposition method, a small quantity of a heavy fluid may form a floating drop with two different equilibrium radially symmetric shapes: one with a smaller and one with a larger radius of the triple phase contact line.  It should be emphasized that the existence of two different equilibrium shapes with a fixed drop volume was first mentioned in \citep{Hartland1976}, but the shape with the smaller value of the contact radius was discarded as being unstable. We demonstrate here theoretically and experimentally that both shapes are, in fact, stable for drop volumes below a certain critical value. 

Our next goal is to experimentally study the dynamic floatability of a vertically vibrated water drop on the surface of a more viscous and less dense oil bath. Contrary to its anticipated destructive effect, external vibration is known to suppress the Rayeigh-Taylor instability in stratified liquid films and in liquid films on the underside of a solid plate \citep{Wolf1969,Wolf1970,Lapuerta2001,Bestehorn2017,BP16,PB16b}. In a series of recent experiments, vertical shaking was shown to create levitating layers of a heavy fluid up to 20 cm in width floating on top of a lighter fluid \citep{Fort_2020}. A rather remarkable phenomenon of the upside-down floatability was revealed, when a solid body positioned at the lower interface of a levitating layer, acquired a stable buoyancy position under the action of vertical vibration \citep{Fort_2020}.

Earlier, we developed a long-wave hydrodynamic model to investigate the saturation of the Rayleigh-Taylor instability in isolated vertically vibrated two-dimensional liquid drops on the surface of a finite-thickness carrier liquid film \citep{POB19}. In the absence of vibration, a small quantity of a heavy fluid, deposited on the surface of a less dense carrier film stretches into a liquid column. The tip of the extending column eventually reaches the solid substrate and the carrier film ruptures. We have shown that an external vertical vibration prevents film rupture at non-zero Reynolds numbers, leading to a formation of a stable floating drop.  

Motivated by these findings, we conduct here a series of experiments with water drops on a vibrated oil surface and demonstrate that in a certain range of the vibration amplitudes and frequencies, the floatability of the drop is enhanced, allowing a larger quantity of water to stay afloat. In this regime, the drop elongates horizontally driven by the subharmonic surface Faraday waves that develop in its upper sessile cap \citep{Pucci_2011,Pucci_2013}. The time-averaged total length of the triple-phase contact line increases for the elongated drop, leading to a larger lifting force generated by tensile forces.  This new dynamical regime exists in a narrow window of the vibration amplitudes: above the onset of the Faraday waves in the drop and below the onset of the Faraday waves on the more viscous oil surface. An even stronger vibration destroys the balance of vertical forces and the drop sinks.

The paper is organized as follows. Section\,\ref{stat} presents
the summary of the analysis of equilibrium drop shapes  in the static system in case when the line tension is neglected. The set of the minimal surface equations for each of the three interfaces is written in the form of a boundary-value problem with integral 
constraints that take into account the vertical force balance and satisfy the Neumann triangle condition at the triple-phase contact line. We use the method of numerical continuation implemented with AUTO \citep{AUTO} to continue the analytically known solution that corresponds to a spherical drop in case of zero gravity towards non-zero values of $g$. For any fixed drop volume at sufficiently small gravity $g$, there exist two different equilibrium shapes: one with a larger and one with a smaller radius of the circular triple phase contact line. In Section\,\ref{stability}, we use the Helmzoltz free energy to study the static stability of the drop and find that both equilibrium shapes can be stable for sufficiently small drops. This conclusion is experimentally verified using a small $< 10$ $\mu$L water drop deposited onto the surface of a commercial vegetable oil. The effect of the vertical vibration is experimentally studied in  Section\,\ref{vibr}.

\section{Stationary floating drop}
\label{stat}
\begin{figure}
\centerline{\includegraphics[width=0.99\hsize]{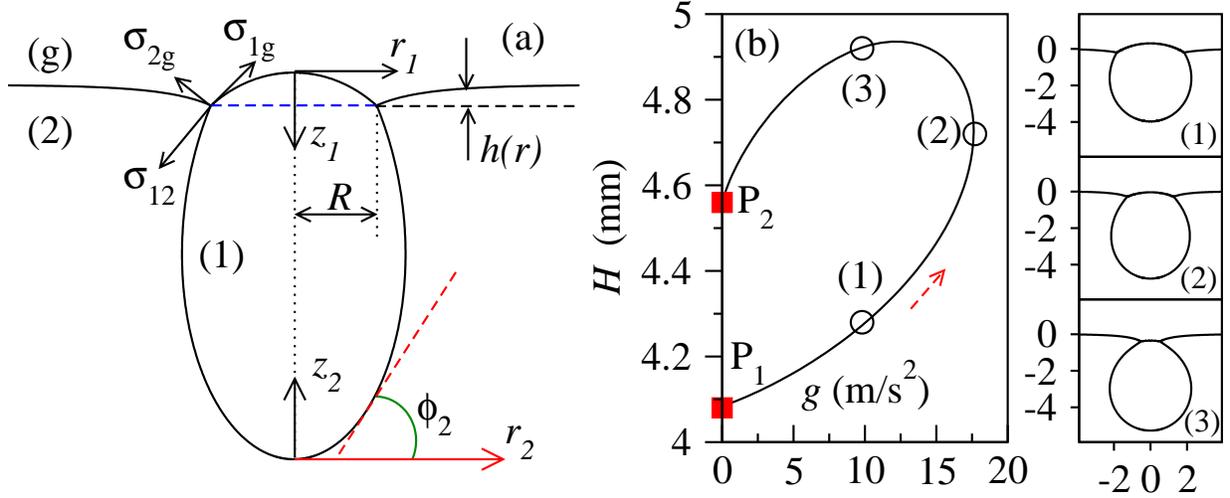}  }
\caption{\label{F1} (a) A liquid drop of a heavy fluid (1) floating at the interface between a lighter fluid (2) and gas (g). Equilibrium Newmann's triangle at the  triple-phase contact line is the requirement that the net force acting on a small element of the contact line is zero, i.e. ${\bm \sigma}_{12}+{\bm \sigma}_{1g}+{\bm \sigma}_{2g}=0$. (b) Variation in
the total drop height $H$ on the gravity constant $g$ for a $50$ $\mu L$ water drop floating at the oil-air interface.  The fluid
parameters are:
$\rho_1=1000$ kg/m$^3$, $\rho_2=916$ kg/m$^3$, $\sigma_{1g}=0.045$ N/m, $\sigma_{2g}=0.022$ N/m, $\sigma_{12}=0.032$ N/m \citep{Phan2012}. 
 The drop profiles at points $(1,2,3)$ are shown in the right panel. Continuation starts at point $P_1$ that corresponds to the analytical solution given by Eqs.\,(\ref{eq12}). The solution at the point $P_2$ corresponds to a circular drop fully submerged in fluid $(2)$ and touching the interface $2-g$ at a point.
}
\end{figure}

We consider a sufficiently small drop of a heavy fluid $(1)$ capable of floating at the interface between a lighter fluid $(2)$, and the ambient gas $(g)$ as shown in Fig.\ref{F1}(a). In equilibrium, the drop is radially symmetric. The shapes of the three interfaces that connect at the triple-phase contact line are found from the balance between the Laplace and hydrostatic pressure. A recent overview of basic phenomena related to  floating droplets referred to as liquid lenses at liquid-gas interfaces, including wetting, dewetting and hydrodynamic instabilities can be found in \citep{NEPOMNYASHCHY2021}. The upper part of the drop, above the horizontal dashed line represents a sessile drop. Using standard set of coordinates \citep{Lohnstein1906,Boucher1975}, the upper tip of the drop is taken as the origin of the local coordinate system, with the $z_1$-axis pointing downward. Introducing an angle $\phi_1$ between the tangent to the drop profile at point $(r_1,z_1)$ and the horizontal, the pressure balance yields \citep{Lohnstein1906,Boucher1975}
\begin{eqnarray}
\label{eq1}
\sigma_{1g}\left(\frac{d \phi_1}{d s_1} +\frac{\sin(\phi_1)}{r_1}\right)=\rho_1 g z_1 +\frac{2\sigma_{1g}}{R_1},
\end{eqnarray}
where $s_1$ is the arc length of  the drop profile from the origin and $R_1$ is the radius of curvature at the upper tip.
Note that the gas density is neglected in Eqs. (\ref{eq1}) and
(\ref{eq2}) below.

Similar, the lower part of the drop, below the horizontal dashed line 
represents a pendant drop. The lower tip of the drop is taken as the origin of the local coordinate system, with the $z_2$-axis pointing upward. At point $\left(r_2,z_2\right)$, $\phi_2$ is the angle between the tangent to the drop profile and the horizontal, and $s_2$ is the arc length measured from the lower tip. Similar to Eq.(\ref{eq1}), the pressure balance can be written as \citep{Boucher1975}
\begin{eqnarray}
\label{eq2}
\sigma_{12}\left(\frac{d \phi_2}{d s_2} +\frac{\sin(\phi_2)}{r_2}\right)=(\rho_2-\rho_1) g z_2 +\frac{2\sigma_{12}}{R_2},
\end{eqnarray}
where $R_2$ is the radius of curvature at the lower tip. 
In addition to Eq.(\ref{eq1}) and (\ref{eq2}), 
the coordinates $(r_i,z_i)$, $i=1,2$ satisfy
\begin{eqnarray}
\label{eq3}
\frac{d r_i}{d s_i}=\cos(\phi_i), ~~\frac{d z_i}{d s_i}=\sin(\phi_i).
\end{eqnarray}
Finally, the meniscus around the drop can be best described in terms of the height $h(r)$, as shown in Fig.1(a). The requirement that pressure is constant at any given vertical level, leads to \citep{Landau}
\begin{eqnarray}
\label{eq4}
\sigma_{2g}\left(\frac{h''}{(1+h'^2)^{3/2}} +\frac{h'}{r\sqrt{1+h'^2}}\right)=\rho_2 g h -\rho_2 g h_0,
\end{eqnarray}
where a prime stands for the derivative with respect to $r$, 
$h_0$ denotes the height of the meniscus far away from the drop
and $r$ is the radial coordinate measured from the vertical 
symmetry axis of the system. 
The system of Eqs.\,(\ref{eq1}) - 
(\ref{eq4}) must be solved as a boundary value problem, supplemented with the following boundary and integral constraints. 

Equations \,(\ref{eq1}) and (\ref{eq3}) with $i=1$ are solved on the interval $0\leq s_1\leq S_1$, where $S_1$ is the arc length of the upper part of the drop measured from the upper tip to the triple-phase contact line. Similar, Eq.\,(\ref{eq2}) and Eq.\,(\ref{eq3}) with $i=2$ are solved on the interval $0\leq s_2\leq S_2$, where $S_2$ is the arc length of the lower part of the drop measured from the lower tip to the  triple-phase contact line. 
At the tips of the drop, we set the initial conditions
\begin{eqnarray}
\label{eq5}
r_i(0)=z_i(0)=\phi_i(0)=0, \,\,i=1,2.
\end{eqnarray}
The requirement of continuity of the interface yields
\begin{eqnarray}
\label{eq6}
r_1(S_1)=r_2(S_2)=R, ~~{\rm {and~~}} z_1(S_1)=z_2(S_2),
\end{eqnarray}
where $R$ denotes the contact radius at the  triple-phase
contact line.
Equation \,(\ref{eq4}) is solved on the interval $R\leq r \leq \infty$ with $h(R)=0$, $h'(\infty)=0$ and $h(\infty)=h_0$.

The Newmann triangle, formed by three tensile forces ${\bm \sigma}_{12}+{\bm \sigma}_{1g}+{\bm \sigma}_{2g}=0$, as shown in Fig.\ref{F1}(a), reflects the requirement that the net force acting on a small element of the contact line vanishes in equilibrium. Introducing the contact angles $\Phi_i=\phi_i(S_i),~i=1,2$ between phase $(i)$ and the horizontal at the triple-phase contact line, Neumann's triangle is replaced by two scalar equations
\begin{eqnarray}
\label{eq7}
0&=&\sigma_{1g}\cos{(\Phi_1)}+\sigma_{12}\cos{(\Phi_2)}-\sigma_{2g}\frac{1}{\sqrt{1+\left[h'(R)\right]^2}},\nonumber\\
0&=&\sigma_{1g}\sin{(\Phi_1)}-\sigma_{12}\sin{(\Phi_2)}+\sigma_{2g}\frac{h'(R)}{\sqrt{1+\left[h'(R)\right]^2}}.
\end{eqnarray}

The volumes $V_i,~ i=1,2$ of the sessile and pendant parts are given by
\begin{eqnarray}
\label{eq8}
V_i=\int_0^{S_i}  \pi r_i^2\sin(\phi_i)ds_i.
\end{eqnarray}
The total buoyancy force $F_b$ experienced by the drop is given by the weight of fluid $2$ displaced by the volume of pendant part $\rho_2 g V_2$ and the weight of fluid $2$ in the cylindrical volume above the contact line level ( the dashed line in Fig.\,\ref{F1}(a)) with  the radius $R$ and the meniscus height $h_0$, i.e. 
\begin{eqnarray}
\label{eq9}
F_b=\rho_2 g V_2+\rho_2 g \pi R^2 h_0.
\end{eqnarray}

Finally, the  vertical force balance implies that the weight of the drop $\rho_1 g (V_1+V_2)$ is balanced by the buoyancy force $F_b$ and the vertical component of the surface tension force exerted by fluid $(2)$ onto the drop. The latter is given by $ \displaystyle 2 \pi \sigma_{2g} R h'(R)\left[1+\left[h'(R)\right]^2\right]^{-1/2}$ so that the balance of vertical forces can be written in the form
\begin{eqnarray}
\label{eq10}
\rho_1 g (V_1+V_2)=F_b+2 \pi \sigma_{2g} R \frac{h'(R)}{\sqrt{1+\left[h'(R)\right]^2}}.
\end{eqnarray}
 
We use numerical continuation method AUTO \citep{AUTO} to solve the above boundary value problem with an integral condition given by the requirement that the total volume $V_1+V_2$ is fixed. Details of the numerical continuation method are summarized in Appendix\,\ref{appA}.

As a starting point of the numerical continuation method, we use the analytical solution in case of zero gravity, when the upper and the lower parts of the drop are both spherical and the interface between fluid $2$ and gas is non-deformed. By setting $h'(R)=0$, we obtain from Eqs.\,(\ref{eq7}) the contact angles $\Phi_i$ 
\begin{eqnarray}
\label{eq11}
\Phi_1&=&\cos^{-1}\left(\frac{\sigma_{1g}^2+\sigma_{2g}^2-\sigma_{12}^2}{2\sigma_{1g}\sigma_{2g}}\right),\nonumber\\
\Phi_2&=&\cos^{-1}\left(\frac{\sigma_{12}^2+\sigma_{2g}^2-\sigma_{1g}^2}{2\sigma_{12}\sigma_{2g}}\right).
\end{eqnarray}
It is easy to see that the solution of  Eqs.\,(\ref{eq1}) - (\ref{eq3}) 
at zero gravity is
\begin{eqnarray}
\label{eq12}
\phi_i=\frac{s_i}{R_i},~~r_i=R_i \sin\left(\frac{s_i}{R_i}\right),~~z_i=R_i\left[1-\cos\left(\frac{s_i}{R_i}\right)\right],~~0\leq s_i \leq R_i \Phi_i,
\end{eqnarray}
where the radii $R_i$ of the sessile and pendant parts are determined by the contact angles $\Phi_i$ and the horizontal radius $R$ at the level of the contact line, i.e. $R_i\sin(\Phi_i)=R$.
The volumes of the sessile and pendant parts are found from Eq.\,(\ref{eq8}) $V_i=\pi R_i^3\left( \frac{2}{3}+\frac13 \cos^3(\Phi_i)-\cos(\Phi_i)\right)$.

First, we continue the solution Eqs.\,(\ref{eq12}) in parameter $g$ with the fluid parameters taken from \citep{Phan2012}. The total height of the drop $H=z_1(S_1)+z_2(S_2)$ is displayed as a function of $g$ in Fig.\ref{F1}b. The continuation starts at the point $P_1$ that corresponds to the analytical solution Eqs.\,(\ref{eq12}). As $g$ increases, the drop profile changes, as shown by the three selected solutions $1,2,3$ in the right panel  Fig. \ref{F1}. No stationary solutions exist past the saddle-node bifurcation point (solution $2$) in terms of $g$. For any $g$ below the saddle-node point, there exist two stationary drop profiles: one with a lower value of the radius $R$ (solution $1$) and one with a larger value of $R$ (solution $3$). At vanishingly small $g$, the second solution (point $P_2$) collapses to a spherical 
drop fully submerged in fluid $2$ that only touches the interface $2-g$ at one point.

\begin{figure}
\centerline{\includegraphics[width=0.99\hsize]{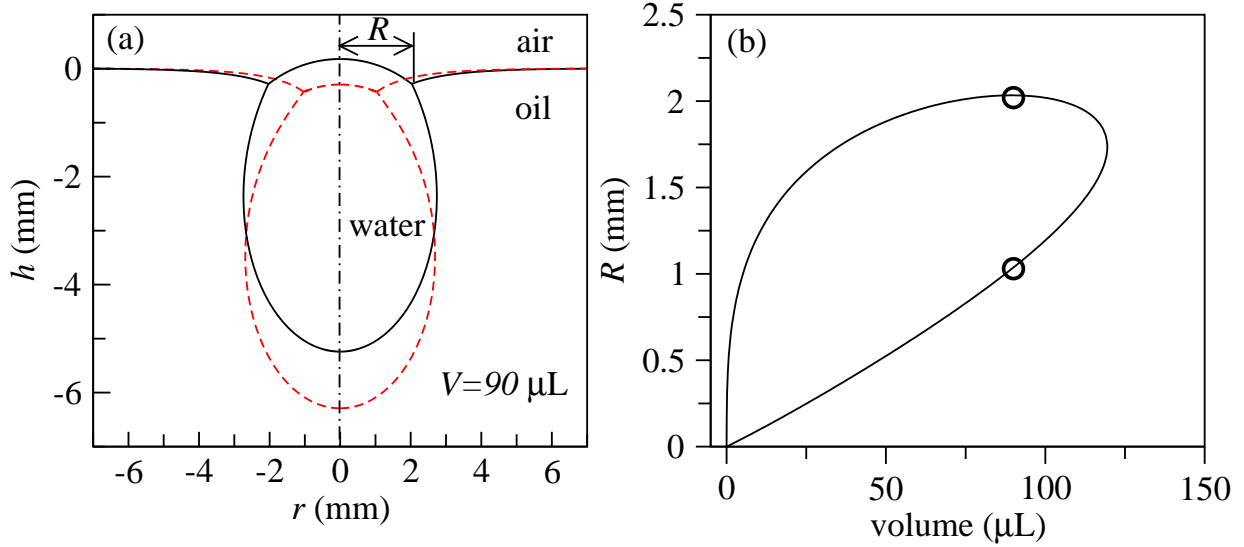}  }
\caption{\label{F2}  (a) Two water drops with the identical volume $V=90$ $\mu$L at the oil-air interface. (b) Contact line radius $R$ as a function of the volume $V$ at $g=9.8$ m/s$^2$. Circles indicate the location of the solutions in (a).
}
\end{figure} 

After the value of $g=9.8$ m/s$^2$ is reached, the solution is followed using
AUTO \citep{AUTO} with the drop volume $V$ as a continuation parameter. The radius $R$ of the contact line is shown as a function of $V$ in Fig.\ref{F2}(b). Two different drops with  the identical volume $V=90$ $\mu$L are shown in Fig.\ref{F2}(a).

It is remarkable that multiple stationary profiles of floating drops have never been studied in details. In the early study \citep{Hartland1976}, the existence of multiple profiles has been mentioned, but the solutions with the smaller value of $R$ were declared 
unstable. The reason for the instability \citep{Hartland1976} was the qualitative comparison of a floating drop with a heavy rigid sphere capable of floating on the surface of a fluid that does not wet the surface of the sphere \citep{Hartland1971}. Indeed, in case of a rigid sphere, the stability condition is derived by looking at the position of the center of mass $z_c(g)$ of the sphere as a function of the gravity constant $g$. Thus, if $g$ is slightly increased (decreased), the center of mass of a stable configuration must descend (ascend) 
so that after $g$ is set back to its original value, the sphere will bounce back up (down) to recover the original stationary position. It was shown \citep{Hartland1971} that solutions with a larger radius of the contact line are stable, while those 
with the smaller value of radius are unstable. However, to the best of our knowledge, no such calculations are available for a floating liquid drop. A comprehensive stability analysis that takes into account static and dynamic perturbations is still missing.

\section{Static stability of a floating drop}
\label{stability}
We approach the stability of a floating drop from the point of view of the Helmholtz free energy, by considering the so-called static axisymmetric perturbations of the drop profile that satisfy the Laplace pressure balance. Such static stability analysis neglects the motion of the fluids and, therefore, can only be considered as a precursor to the true dynamic stability with respect to non-axisymmetric perturbations. Note that a similar method has been applied to study the static stability of pendant drops hanging from a solid plate \citep{Padday1973}.

\begin{figure}
\centerline{\includegraphics[width=0.5\hsize]{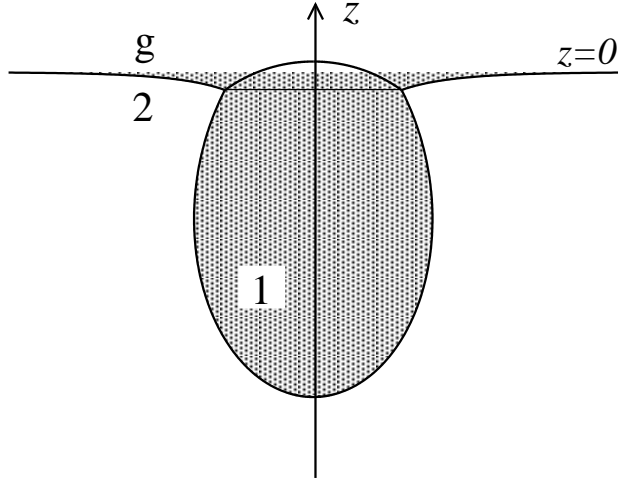}  }
\caption{\label{F3}  Excess potential energy $U_e$ of the fluid bath is given by the potential energy of the shaded part filled with fluid $2$ with negative density $-\rho_2$.
}
\end{figure}

We take the level of the carrier fluid $2$ far away from the drop, as a zero level for the vertical axis directed upwards, as shown in Fig.\ref{F3}. The excess total energy $E_t$ of the system is constructed not counting the infinite energy of the semi-infinite fluid $2$ with a flat fluid-gas interface in the absence of the drop. 

The excess surface energy is 
\begin{eqnarray}
\label{sen}
E_s=\sigma_{12} S_{12} +\sigma_{1g}S_{1g}+\sigma_{2g} S_{2g},
\end{eqnarray}
where $S_{12}=\int_0^{S_2} 2\pi r_2ds_2$ and $S_{1g}=\int_0^{S_1} 2\pi r_1ds_1$ are the surface areas of the fluid\,$1$-fluid\,$2$ and fluid\,$1$-gas interfaces and $S_{2g}=\int_R^\infty 2\pi r(\sqrt{1+h'^2}-1)dr$ is the excess surface area of the fluid\,$2$-gas interface.
 The excess potential energy of the system is
\begin{eqnarray}
\label{pen}
E_p=Mgz_c +U_e,
\end{eqnarray}
where $M=\rho_1 (V_1+V_2)$ is the total mass of the drop, $z_c$ is the vertical coordinate of the centre of mass of the drop and $U_e$ is the excess potential energy of the fluid bath. $U_e$ can be further represented in the form
\begin{eqnarray}
\label{ue}
U_e=-g\rho_2 V_e Z_e,
\end{eqnarray}
 where $V_e$ is the volume of the shaded part in Fig.\ref{F3} and $Z_e$ is the coordinate of the center of mass of the shaded part in Fig.\ref{F3}, filled with a homogeneous density.
The excess total energy $E_t$ is thus given by $E_t=E_p+E_s$.
 Note that the values $z_c,Z_e,V_e$ and $E_s$ depend on $g$, thus
\begin{equation}
z_c=z_c(g),~Z_e=Z_e(g),~V_e=V_e(g),~E_s=E_s(g).
\end{equation}

Now let us assume that the gravity constant has been slightly changed to $g+\delta g$. This can be practically achieved by placing the system in a lift that accelerates either upward or downward, or alternatively, as a result of a localized in time displacing disturbance of the system. The drop will assume a new stationary shape that corresponds to the new value of $g+\delta g$. Now assume that the lift (or the disturbance) suddenly stops and the gravity constant is instantaneously set back to the original value $g$.  At this moment of time, the drop and meniscus still have  new shapes, so that the total energy of the system is
\begin{eqnarray}
\label{en_new}
\tilde{E}_t= Mgz_c(g+\delta g) - g\rho_2 V_e(g+\delta g) Z_e(g+\delta g)+E_s(g+\delta g).
\end{eqnarray}
The system is stable if $\tilde{E}_t > E_t$, i.e.
\begin{eqnarray}
\label{en_delta}
\tilde{E}_t-E_t&=& Mgz_c(g+\delta g)-Mgz_c(g) \nonumber\\
&-& g\rho_2 V_e(g+\delta g)Z_e(g+\delta g)+g\rho_2 V_e(g)Z_e(g)\nonumber\\
&+&E_s(g+\delta g)-E_s(g)>0.
\end{eqnarray}
Expanding Eq.\,(\ref{en_delta}) into powers of $\delta g$, we obtain the leading term
\begin{eqnarray}
\label{en_delta1}
\delta E_t = \left( Mg\frac{d z_c(g)}{dg}  - g\rho_2 \frac{d(V_eZ_e)}{dg}+\frac{dE_s(g)}{dg} \right)\delta g=0,
\end{eqnarray}
which must vanish for arbitrary small $\delta g$, because the drop shape at the original value of $g$ is in equilibrium.

Differentiating $\delta E_t/\delta g=0$ with respect to $g$, we obtain
\begin{eqnarray}
\label{en_diff}
\frac{d}{dg}\left(\frac{\delta E_t}{\delta g}\right) = M\frac{d z_c(g)}{dg}+Mg \frac{d^2 z_c(g)}{dg^2} - \rho_2 \frac{d(V_eZ_e)}{dg} -g\rho_2 \frac{d^2(V_gZ_g)}{dg^2}+\frac{d^2E_s(g)}{dg^2}=0.~~
\end{eqnarray}
The second term in the expansion of Eq.\,(\ref{en_delta}) in terms of $\delta g$ is
\begin{eqnarray}
\label{en_delta2}
\delta^2 E_t = \left(Mg\frac{d^2z_c(g)}{dg^2} - g\rho_2 \frac{d^2(V_eZ_e)}{dg^2}+\frac{d^2E_s(g)}{dg^2}\right)\delta g^2 >0.
\end{eqnarray}
Equation \,(\ref{en_diff}) can now be used to remove the second derivative terms in Eq.\,(\ref{en_delta2}) and to finally obtain the static stability condition
 \begin{eqnarray}
\label{en_stab}
\rho_2 \frac{d(V_eZ_e)}{dg}-M\frac{d z_c(g)}{dg}>0.
\end{eqnarray} 
Equation \,(\ref{en_stab}) is now recast 
in the form $df(g)/dg<0$, where
 \begin{eqnarray}
\label{en_stab1}
f(g)= z_c(g) -\frac{\rho_2 V_e(g) Z_e(g)}{\rho_1 V},
\end{eqnarray}
and $V$ is the fixed drop volume. Function $f(g)$ can be expressed in terms of the variables $z_i$, $\phi_i$, $r_i$ and $h(r)$ from Eqs.\,(\ref{eq1}) - (\ref{eq3}), as shown in Appendix\,\ref{appB}.


%
\begin{figure}
\centerline{\includegraphics[width=0.9\hsize]{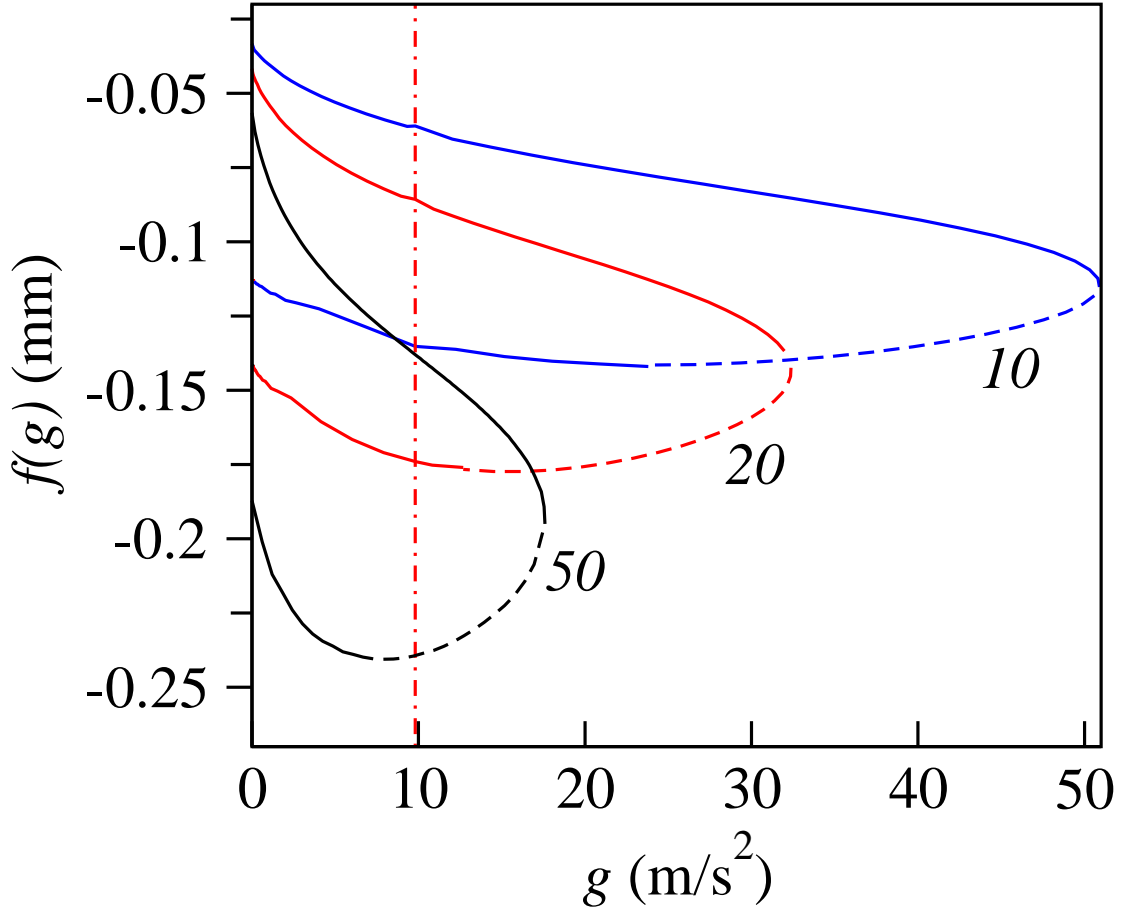}  }
\caption{\label{F4}  Function $f(g)$ given by Eq.\,(\ref{en_stab1}) determined for $ V=50$, $20$ and $10$ $\mu$L water drop floatation on oil with the parameters as in Fig.\ref{F1}(b). Stable (unstable) solutions are marked by the solid (dashed) lines. The zero gravity solution given by Eqs.\,(\ref{eq12}) corresponds to the larger value of $f$ at $g=0$. For each volume, the solutions on the upper branch have a larger contact radius $R$. Drops smaller than $20$ $\mu$L have two stable floatation shapes at $g=9.8$ m/s$^2$: one with a larger and one with a smaller value of the contact radius.
}
\end{figure} 

The function $f(g)$ from Eq.\,(\ref{en_stab1}) is shown in Fig. \ref{F4} for three different volumes $V=50$, $20$ and $10$ $\mu$L of a water drop on oil surface with fluid parameters as in Fig.\ref{F1}(b). Statically stable shapes correspond to $df/dg<0$, as indicated by the solid  curves, whereas statically unstable states are marked by the dashed parts of the curves. For each volume, the zero-gravity solution given by Eqs.\,(\ref{eq12}) has a larger value of $f(g=0)$.  At any fixed value of $g$ below the saddle-node bifurcation point, the solution with larger (smaller) values of the contact radius $R$ belongs to the upper (lower) branch of $f(g)$. At terrestrial gravity of $g=9.8$ m/s$^2$, as indicated by the vertical line in Fig. \ref{F4}, larger drops with the volume $V>~ 20$ $\mu$L have only one stable shape with a larger value of $R$ (upper branch). However, smaller drops with  the volume $V<~ 20$ $\mu$L have two stable shapes: one with a larger (upper branch) and one with a smaller (lower branch) contact radius $R$.



%
\begin{figure}
\centerline{\includegraphics[width=0.9\hsize]{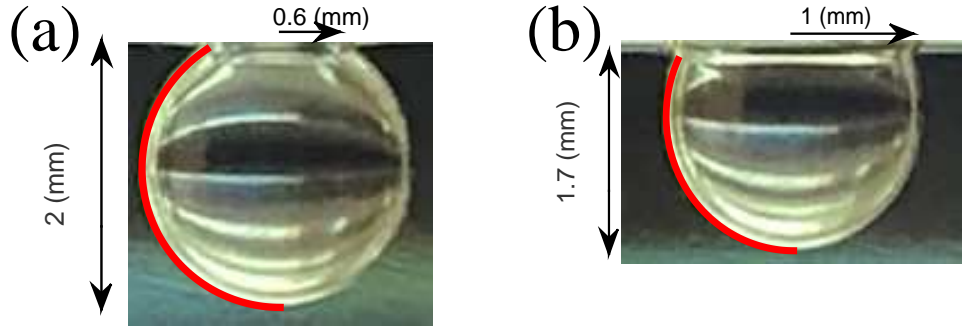}   }
\caption{\label{F5} Two stable shapes of a water drop with approximately identical volume $5\pm 0.5$ $\mu$L floating on the surface of a commercial vegetable oil. The thick solid line is obtained by fitting the experimental drop profile with a pendant drop solution of Eqs.\,(\ref{eq2}), (\ref{eq3}) with the fluid parameters as in Fig.\ref{F1} and the fitted value of the radius of curvature at the lower tip $R_2\approx1.13$ mm in both cases. The shape with a smaller contact radius (a) is obtained by depositing a drop from a pipette directly onto the oil surface. The shape with the larger contact radius (b) is obtained by depositing the drop from the pipette onto the vertical wall of the container and allowing the drop to slowly slide down the wall to touch the oil surface.
}
\end{figure} 

In order to experimentally verify the existence of two stable floatation shapes, we use two different methods to deposit a small quantity of water $<10$ $\mu$L onto a surface of a commercial vegetable oil. To obtain a shape with a smaller contact radius, we use a similar technique as in 
\citep{Phan2012}, namely, a pendant water drop with the maximal possible volume is produced freely hanging in air from a pipette with the diameter of the nostril of $2.5$ mm. After the drop is carefully brought into contact with the oil surface, it separates from the pipette and starts to float on the oil surface, as shown in Fig.\ref{F5}(a). In the second deposition method, we produce a drop with a similar volume hanging freely from a pipette. The drop is carefully brought into contact with a vertical wall of the polystyrene container filled with oil.  The drop slowly slides down the wall under the action of gravity, and eventually comes into contact with the meniscus formed by oil. The spreading coefficient of water on oil is generally larger than that of water on polystyrene. Therefore, the drop is pulled by capillary forces away from the container wall. This process creates the second stable shape with a larger contact radius, as shown in Fig.\ref{F5}(b). In both cases, the estimated volume of the water drop is approximately identical $V\approx 5\pm 0.5$ $\mu$L. 

The drops are photographed and the contours of their pendant parts submerged in oil are extracted using Matlab. Note that the contour of the upper cap of the drop is difficult to extract, as it is obstructed by the meniscus formed by oil with the container wall. The experimentally obtained profiles are fitted by the analytical solution of the pendant drop Eqs.\,(\ref{eq2}) and (\ref{eq3}) with fluid parameters as in Fig.\ref{F1}. We use the radius of curvature at the lower tip $R_2$ as the fitting parameter and obtain the value of $R_2\approx1.13$ mm for both shapes. The analytical solution is shown by the thick solid line in Fig.\ref{F5}. The volumes of the submerged parts, determined using the fitted analytical solution, are found to be $6$ $\mu$L in Fig.\ref{F5}(a) and $5$ $\mu$L in Fig.\ref{F5}(b). The difference in volumes is due to the unaccounted volume of the obstructed upper caps.
\section{Vertical vibration helps liquid drops to stay afloat}
\label{vibr}

\begin{figure}
\centerline{\includegraphics[width=0.99\hsize]{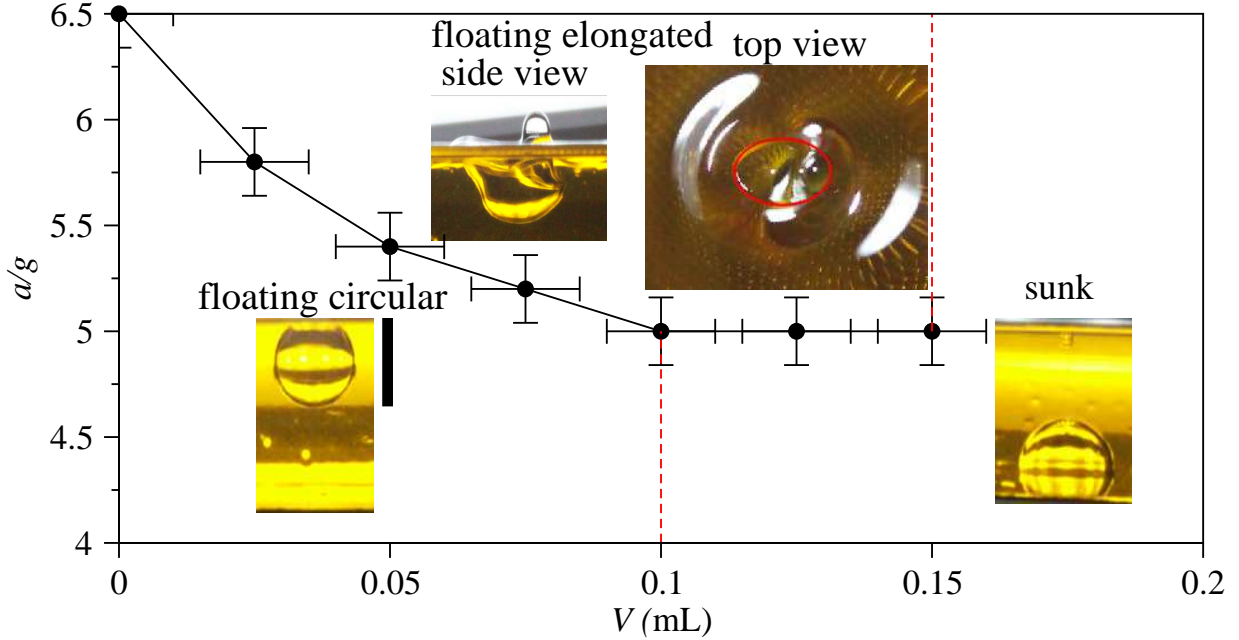}  }
\caption{\label{F6} Phase diagram separating floating circular, floating elongated and sinking water drops vibrated at $60$ Hz in an olive oil. Symbols indicate an experimentally determined critical acceleration $a/g$, that corresponds to the onset of the Faraday instability in a hanging water drop of the varying volume. In the absence of vibration, smaller drops with $V \lesssim 0.1$ mL hang at the oil-air interface, while larger drops $V \gtrsim 0.1$  mL sink to the bottom of the container. For  $0.1$ mL $ \lesssim V \lesssim 0.15$ mL the drop floats at the oil-air interface when vibrated with the amplitude $a/g>5$.  Insets: representative snapshots of a floating circular, floating elongated and a sunk water drop. The vertical scale bar next to the floating circular drop indicates the length of 5 mm. An ellipse (solid line) in the top view of the floating elongated drop highlights an approximate location of the contact line. The ratio between the major and the minor
axes of the ellipse is $2:3$.  
}
\end{figure} 

To investigate how external vertical vibration affects the floatability of the drop, we conduct a series of experiments with drops of distilled water with the volume $V \lesssim 0.15$\,mL deposited on top of a vertically vibrated $1$ cm thick layer of commercial olive oil. The oil was placed in a circular container with 
the diameter of 10 cm mounted on a $6.5$'' $45$\,W RMS audio speaker (Sony,
Japan) powered by a $30$W stereo amplifier (Yamaha TSS-15, China) of a
sinusoidal signal produced by a digital tone
generator. The container was vertically vibrated at frequencies in the range
40--80\,Hz. The vibration amplitudes were sufficiently large to excite
subharmonic Faraday waves on the surface of the water drop but not strong
enough to excite Faraday waves on the free surface of the oil layer.

The vertical coordinate of the container $z_{\rm {p}}(t)$ oscillates with the
amplitude $A$ and the period $T=1/f$ according to
\begin{eqnarray}
\label{Exp_eq1}
z_{\rm p}(t)=A\cos{(\phi+2\pi t/T)},
\end{eqnarray}
where $f$ is the frequency in Hertz and $\phi$ is  a fixed phase.
In the co-moving frame of reference, the time-dependent {\it downward} acceleration is 
\begin{eqnarray}
\label{Exp_eq2}
g+\ddot{z}_{\rm {p}}(t)=g-A\left(\frac{2\pi}{T}\right)^2\cos{(\phi+2\pi t/T)},
\end{eqnarray}
where $\dot{z}$ stands for the time derivative of $z$.

We use the ADXL\,326 accelerometer 
(Analog Devices, USA) attached to the container and a digital
oscilloscope (Tektronix TDS 210, USA) to measure 
 the acceleration amplitude $ \displaystyle \frac{a}{g}=\frac{A(2\pi f)^2}{g}$
in units of $g$.
The motion of the
drop was recorded at up to 240 fps using a high-speed camera and the obtained 
images were post-processed in Matlab for a further analysis.

As a first step, we determine the maximal possible volume of a water drop $V_m$ that hangs at the olive oil-air interface without vibration and find that $V_m \approx 0.1$ mL. Note that with a commercial vegetable oil, a stable hanging water drop of up to $0.17$ mL  can be achieved \citep{Phan2012}. Such a drop stretches down to reach a vertical size of $1$\,cm and more due to a larger value of the vegetable oil-air surface tension. In order to avoid any direct contact between the drop and the bottom of the $1$\,cm deep container used in our experiments, we have chosen olive oil over vegetable oil.

Previous studies have demonstrated that liquid lenses of a lighter fluid floating on the surface of a heavier and more viscous fluid undergo a spontaneous horizontal elongation under the action of vertical vibration \citep{Pucci_2011,Pucci_2013, Pucci_2015,Pototsky_2018}. The underlying physical mechanism was traced down to the Faraday instability at the upper drop surface. The onset of the Faraday instability corresponds to a period-doubling bifurcation \citep{Pototsky_2018}: the lens starts to oscillate at half of the driving frequency and elongates horizontally in a randomly chosen direction. The resultant equilibrium shape of the elongated lense, as seen from the top, is dictated by the balance between the radiation pressure of the unidirectional Faraday waves and the Laplace pressure which always tries to return the lens back into a circular shape \citep{Pucci_2011,Pucci_2013}.

Here we observe that similar to floating lenses, a heavy water drop that floats on the surface of a lighter and more viscous oil also undergoes a horizontal elongation, when Faraday waves are excited on the upper drop surface. Remarkably, the floatability of the elongated drop increases, allowing drops with volumes greater than the maximal static volume $V_m\approx 0.1$ mL to stay afloat.

Our experimental protocol consists of a careful deposition of a water drop of various volumes with a pipette on the surface of the vibrated oil bath, and different acceleration amplitudes $a$. Below the Faraday instability threshold, the drop assumes a circular shape when viewed from the top  as seen in the upper right inset of Fig. \ref{F6}. A side view of such a floating circular drop is also shown in the inset of Fig.\,\ref{F6} (floating circular). Above the Faraday threshold the drop elongates as displayed in the inset in Fig.\,\ref{F6} (floating elongated). The shape of the contact line in the elongated state is approximately elliptic with a relatively small eccentricity  (ratio of the major and minor 
axes) of $2:3$. Contrary to an anticipated destructive effect of shaking, the drop continues to float, when vibrated at up to $a=6.5g$. For $a>6.5g$, the Faraday instability in the oil layer sets in.

Following this protocol we record the Faraday stability threshold for different drop volumes in Fig.\,\ref{F6} vibrated at $f=60$ Hz (a representative value). 
The vertical and horizontal error bars indicate the uncertainty 
in measurements of the acceleration amplitude and the drop volume, respectively. 

We found that drops with the volume $0.1$ mL~<~$V~<~0.15$ mL can float at 
the oil-air interface when vibrated at $a>5g$. These drops will otherwise sink to the bottom of the container when the vibration amplitude is reduced below $a<5g$.
Notably, the action of vibration delays the emergence of the critical event of drop sinkage similar to delaying of film rupture \citep{Bestehorn2017} or delaying of liquid bridge breaking \citep{Benilov2016}.

\subsection{The origin of the excess lifting force}
\label{origin}

In order to understand the physical mechanism responsible for the excess lifting force, required to keep larger drops afloat
we consider the balance of the vertical forces acting on the drop averaged over at least two forcing periods $2T$:
\begin{eqnarray}
\label{Exp_eq4}
Mg= \langle F_b \rangle+\langle F_t \rangle,
\end{eqnarray}
where $M$ is the total mass of the drop, $F_b$ is the buoyancy force, $F_t$ is the lifting force due to surface tension acting at the triple-phase oil-water-air contact line and $\langle \dots \rangle=(2T)^{-1}\int_0^{2T}(\dots )\,dt$ denotes averaging over time. Note that averaging over $2T$ is necessary due to a subharmonic nature of the Faraday waves.

Similar to Eq.\,(\ref{eq10}) in the static case, the instantaneous value of $F_t$ for the elongated drop can be represented in the form
\begin{eqnarray}
\label{Exp_ft}
 F_t =\sigma_{2g}\int_{C(t)} \sin(\alpha(l,t))\,dl,
\end{eqnarray}
where $\alpha(l,t)$ is the local instantaneous contact angle between the oil-gas interface and the horizontal, and the integration contour $C(t)$ represents the instantaneous shape of the contact line of the elongated drop, parametrised by the arc length $l$. In the static case, $ \sin{\left(\alpha(l,t)\right)}=h'(R)/\sqrt{1+\left[h'(R)\right]^2}$ is constant and $C$ is a circle with the
radius $R$ so that Eq.(\ref{eq10}) is recovered.

\begin{figure}
\centerline{\includegraphics[width=0.99\hsize]{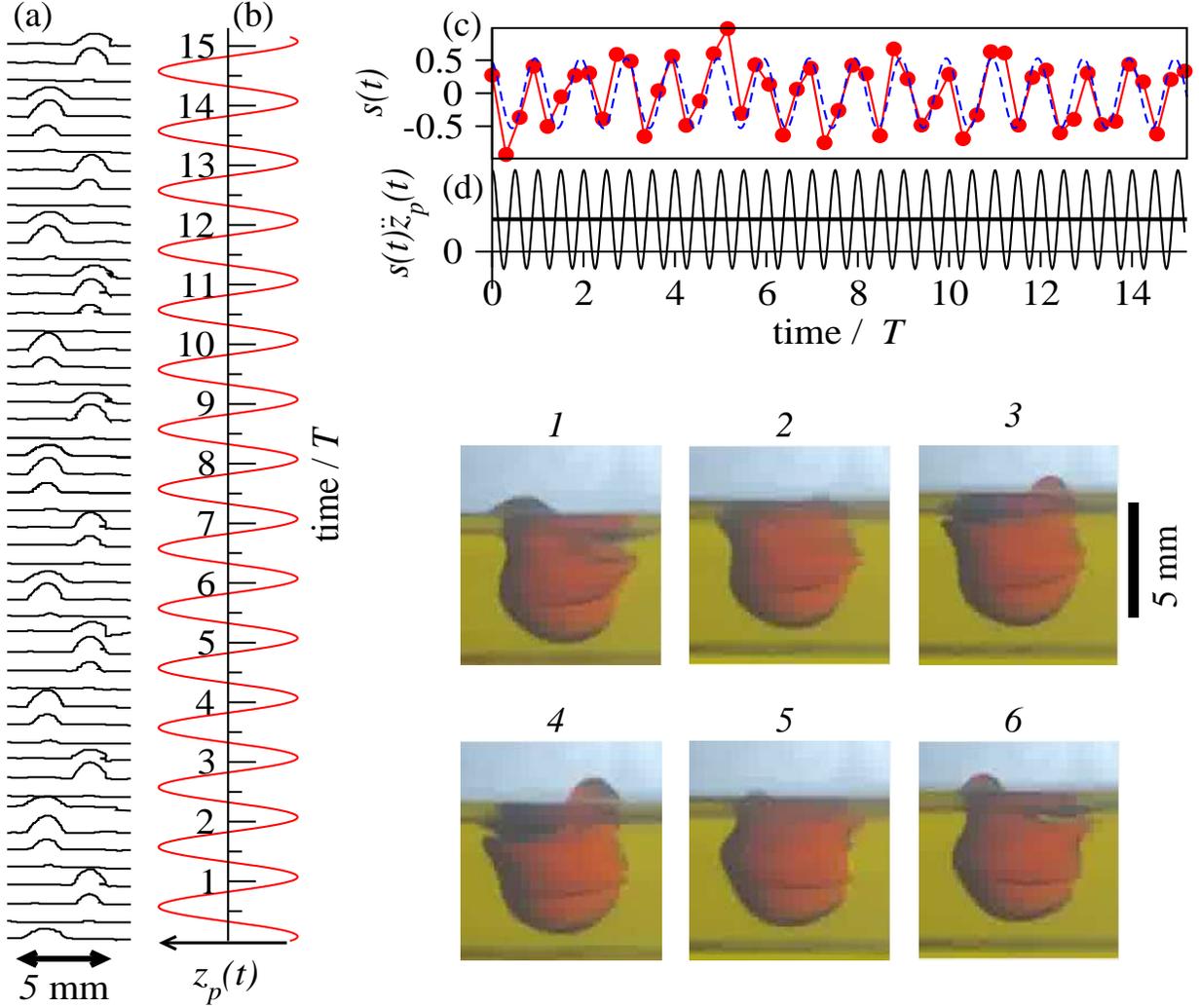} }
\caption{(a) Series of $50$ contours of the upper part of a $0.12$ mL floating drop vibrated at $60$ Hz extracted from a 200 fps slow motion video. The vertical axes in (a,b) represent time in the units of the vibration period $T=1/60$ s. The horizontal axis in (b) is the vertical coordinate of the container $z_{\rm {p}}(t)$ vibrating at $60$ Hz. (c) Circles correspond to the scaled volume factor $s(t)$ associated with the upper part of the drop above the oil-air interface estimated  
from (a). The dashed curve in (c) is the fitting line 
$\sim S_0\cos(\psi+ 2\pi t/T)$ to the experimental data. (d) The function $s(t)\ddot{z}_{\rm {p}}(t)$ together with its time-average value represented by the thick horizontal line. 
Panels 1 - 6 display snapshots of the drop showing also the blobs emerging above the oil-gas interface. The six snapshots correspond to the first six contours in (a) from the time interval between $0$ and $1.5T$. } 
\label{F7}
\end{figure}

The excess lifting force can be explained if $\langle F_t \rangle$ is larger than its static value, given by $2\pi R\sigma_{2g}h'(R)/\sqrt{1+\left[h'(R)\right]^2}$. However, it is important to emphasize that the average value of the integral in Eq.\,(\ref{Exp_ft}) cannot be simply estimated, due to the time-dependent contact angle $\alpha(t,l)$, as can be appreciated from the side view of the floating elongated drop in the inset in Fig.\ref{F6}.

As we explain below, the estimation of the average value of the buoyancy force $ \langle F_b \rangle$ is experimentally easier accessible than the estimation of $\langle F_t \rangle$. We therefore proceed to show that $ \langle F_b \rangle$ measured for the vibrated drop is, in fact, smaller than the static value of a fully submerged drop, i.e. $\langle F_b \rangle<(\rho_1-\rho_2)gV_0$, where $V_0$ represents the total volume of the drop. This confirms that the excess lifting force is due to the increased value of $\langle F_t \rangle$.

We record a side-view slow-motion video of a $0.12$ mL drop. To increase contrast with yellow oil, we add a small amount of red food coloring fluid to the water. The drop is stretched horizontally and the stretching direction is perpendicular to the view line of the camera, as shown in the inset in Fig.\ref{F6}. 

Individual frames are extracted from the slow-motion video and post-processed in Matlab to detect the contour of the drop. Figure \,\ref{F7}(a) displays $50$ contours of the upper part of the drop exposed 
above the oil-air interface. The time interval between any two neighboring contours is $1/200$ s. For comparison, we show in Fig.\ref{F7}(b) the vertical coordinate of the vibrating container $z_c(t)$ in the laboratory frame. It is clear that the drop oscillates sub-harmonically at half of the driving frequency $f/2=30$ Hz. The left and the right edges of the drop periodically bulge out to develop a small blob every $1/30$ s in anti-phase to each other.

Before proceeding, we estimate the average dynamic pressure $P_d$ acting on the water-oil interface at the submerged part of the drop. Due to continuity of the velocity field, the difference of the dynamic pressures in oil and water is given by $P_d= |\rho_1-\rho_2| u^2$, where $\rho_1$ and $\rho_2$ 
are the densities of water and oil, respectively.  The typical velocity inside the water drop can be extracted from Fig.\ref{F7}(a). 
The maximal height of the blobs periodically developing above the oil-air interface are $\sim 1\dots3$ mm. Since the blobs develop over the interval of time $1/f=1/60$ s, the average speed associated with this motion is $u\sim 10^{-1}$ m/s. This yields the average dynamic pressure at the oil-water interface of  $P_d \sim |\rho_1-\rho_2| \times 10^{-2}$ N/m$^2$. 
The dynamic pressure must be compared with the hydrostatic pressure 
$P_s=|\rho_1-\rho_2| g h$, 
where $h\approx 10^{-2}$ m is the vertical size of the drop. Thus, we estimate that the dynamic pressure $P_d \approx 10^{-1} P_s$ can be neglected and the total buoyancy force is given by
\begin{eqnarray}
\label{Exp_eq5}
\langle F_b \rangle=(\rho_1-\rho_2) \langle (g+\ddot{z}_{\rm {p}}(t))V_s(t) \rangle,
\end{eqnarray}
where $V_s(t)$ is the time-dependent volume of the oil displaced by the drop.

We assume $V_s(t) \approx V_0 -  \Delta V(t)$, where $V_0$ is the total volume of the drop and  $\Delta V(t)\geq 0$ is the instantaneous volume of the part of the drop above the oil-air interface. 
It is important to note that this assumption neglects the existence of the air pockets (dark regions in the insets in Fig.\ref{F6}). 
Finally, we obtain
\begin{eqnarray}
\label{Exp_eq6}
\langle F_b \rangle=(\rho_1-\rho_2)(gV_0-\langle\ddot{z}_{\rm {p}}(t) \Delta V(t) \rangle-g\langle\Delta V(t)  \rangle ),
\end{eqnarray}
where it was used that $\langle\ddot{z}_{\rm {p}}(t)  \rangle=0$. 
Equation \,(\ref{Exp_eq6}) shows that the average buoyancy force $\langle F_b \rangle$ is larger than the static buoyancy $(\rho_1-\rho_2)gV_0$ only  
if $\langle\ddot{z}_{\rm {p}}(t) \Delta V(t) \rangle + g\langle\Delta V(t)  \rangle< 0$.

Note that the average of a product of two sinusoidal 
functions $\langle f(t)g(t+\phi) \rangle$ 
with a zero average value 
$\langle f(t)\rangle=\langle g(t)\rangle=0$ can be positive or negative 
depending on the phase shift between them. 
For example, $\langle a\cos(\omega t )\cos(\phi + \omega t) \rangle = a\cos(\phi)/2$.

Our analysis is continued to show that 
$\langle\ddot{z}_{\rm {p}}(t) \Delta V(t) \rangle >0 $, which together with $\langle\Delta V(t)  \rangle > 0$ implies that the average buoyancy force is smaller than the static buoyancy 
$\langle F_b \rangle<(\rho_1-\rho_2)gV_0$. 
To show that $\langle\ddot{z}_{\rm {p}}(t) \Delta V(t) \rangle >0 $, we assume that the true volume $\Delta V(t)$ of the upper part of the drop is proportional to the area $S(t)$ under the two-dimensional drop contour, shown in Fig.\,\ref{F7}(a). 
The scaled normalized area $s(t)=S(t)/\langle S(t)\rangle-1 $ is calculated for each of the 50 profiles in Fig.\,\ref{F7}(a) 
and presented versus time in Fig.\,\ref{F7}(c). 
The data is fitted by $S_0\cos(\psi+2\pi t/T)$ shown by the dashed line in 
 Fig.\,\ref{F7}(c). 
The function $\ddot{z}_{\rm {p}}(t) s(t) $ is also shown in Fig.\,\ref{F7}(c). 
along with the time-average value represented by the thick horizontal line. Since $\langle \ddot{z}_{\rm {p}}(t) s(t) \rangle >0$, we finally conclude that $\langle\ddot{z}_{\rm {p}}(t) \Delta V(t) \rangle >0 $. 

Consequently, the floatation of the drop is necessarily enabled by the increased surface tension force $\langle F_t \rangle$, which supports the drop at the triple-phase contact line. The average $\langle F_t \rangle$ becomes larger than the corresponding force in the static case, due to the increased total length of the contact line in a horizontally elongated drop, as observed in our experiments.

\section{Conclusion}
\label{conc}
To conclude, we have studied theoretically and experimentally the stability and dynamics of a liquid drop of a heavier fluid, floating on the surface of a lighter and more viscous fluid. In equilibrium, small drops may stay afloat assuming two different radially symmetric shapes. One stable shape has a smaller and one has a larger value of the triple-phase contact-line radius. We have experimentally demonstrated the possibility to create both stable shapes using two different deposition methods of the water drop onto the oil surface. 
As the volume of the drop is gradually increased beyond a certain critical volume, the shape with a smaller contact radius loses its stability and the drop sinks. The second stable shape with a larger contact radius remains stable until another critical volume is reached, beyond which no {\it static} floating drops exist.  

Remarkably, the floatability of the drop can be slightly increased if the drop is vibrated vertically with the frequency in Hz-order. We have performed a series of experiments with water drops on an olive oil surface and found that drops with the volume between $0.1$ mL and $0.15$ mL remain afloat when vibrated at $60$ Hz with the acceleration amplitude between $5g$ and $6g$. In the absence of vibration, water drops larger than $0.1$ mL detach from the oil surface and sink.  The origin of the excess lifting force is rooted in the horizontal elongation of the drop, driven by the unidirectional Faraday waves that develop on the upper drop surface. The average length of the contact line in an elongated state appears to be larger than in a static radially symmetric shape. As  a result, the average lifting force exerted by tensile forces increases, allowing for heavier drops to stay afloat when vibrated. 

Finally, the results presented here, apart of their academic interest, 
may serve as a basis for a research
in emulsification of dispersed systems containing several 
immissible liquids of different densities and utilized in pharmaceutical
applications for stabilization of emulsions in their suspended state by 
shaking. 

\section*{Acknowledgments}

\appendix
\section{}
\label{appA}
In order to use the numerical continuation method \citep{AUTO}, we write the minimal surface equations Eqs.\,(\ref{eq1})- 
(\ref{eq3}) as a system of nine first-order autonomous equations
\begin{eqnarray}
\label{Aeq1}
\frac{d \phi_1}{d s_1} &=&-\frac{\sin(\phi_1)}{r_1}+\frac{\rho_1 g}{\sigma_{1g}} z_1 +\frac{2}{R_1}, \nonumber\\
\frac{d r_1}{d s_1}&=&\cos(\phi_1),\nonumber\\
\frac{d z_1}{d s_1}&=&\sin(\phi_1),\nonumber\\
\frac{d \phi_2}{d s_2} &=&-\frac{\sin(\phi_2)}{r_2}+\frac{(\rho_2-\rho_1) g}{\sigma_{12}} z_2 +\frac{2}{R_2}, \nonumber\\
\frac{d r_2}{d s_2}&=&\cos(\phi_2),\nonumber\\
\frac{d z_2}{d s_2}&=&\sin(\phi_2),\nonumber\\
\frac{dh}{dr}&=&h',\nonumber\\
\frac{dh'}{dr}&=&(1+h'^2)^{3/2}\left(\frac{\rho_2 g}{\sigma_{2g}} h -\frac{\rho_2 g}{\sigma_{2g}} h_0-\frac{h'}{\tilde{r}\sqrt{1+h'^2}}\right),\nonumber\\
\frac{d\tilde{r}}{dr}&=&1,
\end{eqnarray}
The new variable $\tilde{r}=r$  is introduced to allow writing 
the meniscus equation as an autonomous system of first-order 
differential equations,
where we have taken a special care of Eq.\,(\ref{eq4}) by truncating 
the semi-infinite interval $R\leq r <\infty$ to 
$R\leq r <$ $ R_\infty $, with a fixed upper bound of 
$R_\infty =10$ cm. 
Equations (\ref{eq1})-(\ref{eq3}) are solved on the interval 
$0\leq s_1 \leq S_1$ and Eqs. (\ref{eq4})-(\ref{eq6}) are solved on the interval $0\leq s_2 \leq S_2$.
Note that the first and the fourth equations in Eqs.\,(\ref{Aeq1}) are singular at $s_i=0$ due to the presence of the term $\displaystyle \frac{\sin(\phi_i)}{r_i}$. In order to avoid singularity we introduce the following set of fourteen regularized boundary conditions
\begin{eqnarray}
\label{Aeq2}
\phi_1(0)&=&10^{-5},\nonumber\\
r_1(0)&=&10^{-5}R_1,\nonumber\\
z_1(0)&=&0,\nonumber\\
r_1(S_1)&=&R,\nonumber\\
\phi_2(0)&=&10^{-5},\nonumber\\
r_2(0)&=&10^{-5}R_2,\nonumber\\
z_2(0)&=&0,\nonumber\\
r_2(S_2)&=&R,\nonumber\\
h'(R_\infty)&=&0,\nonumber\\
h(R)&=&0,\nonumber\\
\tilde{r}(R)&=&R,\nonumber\\
0&=&\sigma_{1g}\cos(\phi_1(S_1))+\sigma_{12}\cos(\phi_2(S_2))-\sigma_{2g}\frac{1}{\sqrt{1+\left[h'(R)\right]^2}},\nonumber\\
0&=&\sigma_{1g}\sin(\phi_1(S_1))-\sigma_{12}\sin(\phi_2(S_2))+\sigma_{2g}\frac{h'(R)}{\sqrt{1+\left[h'(R)\right]^2}},\nonumber\\
\rho_1 g (V_1+V_2)&=&\rho_2 g V_2+\rho_2 g \pi R^2 h_0+\sigma_{2g}2\pi R \frac{h'(R)}{\sqrt{1+\left[h'(R)\right]^2}},
\end{eqnarray}
where the last condition represents the balance of the vertical forces acting at the triple-phase contact line. The condition $r_i(0)=10^{-5}R_i$  together with $\phi_1(0)=10^{-5}$ ensure that $\sin(\phi_i)/r_i=R_i^{-1}$ at $s_i=0$.

Finally, we add two integral conditions associated with the volumes of the pendant part $V_2$ and the total drop volume $V_1+V_2$
\begin{eqnarray}
\label{Aeq3}
V_2&=&\int_0^{S_2}\pi r_2^2 \sin(\phi_2)\,ds_2,\nonumber\\
V_1+V_2&=&\int_0^{S_1}\pi r_1^2 \sin(\phi_1)\,ds_1+\int_0^{S_2}\pi r_2^2 \sin(\phi_2)\,ds_2.\nonumber\\
\end{eqnarray}
To be able to continue a solution of the set of nine equations with fourteen boundary and two integral conditions one requires $14+2-9+1=8$ active continuation parameters. These are chosen to be 
$(S_1,S_2,R_1,R_2,R,h_0,V_2)$ with one additional principal continuation parameter, such as, for example, the gravity constant $g$ or the total drop volume $V_1+V_2$. The AUTO files are available on demand.

\section{}
\label{appB}
This section summarizes the computation of the stability function $f(g)$, from Eq.\,(\ref{en_stab1}) in terms of the variables $z_i$, $\phi_i$, $r_i$ and $h(r)$ based on Eqs.\,(\ref{eq1}) - (\ref{eq3}). 
Taking the level of the fluid bath far away from the drop as a zero level, the vertical coordinate of the centre of mass of the drop $z_c$ can be expressed as 
\begin{eqnarray}
\label{azc1}
z_c=\frac{\pi}{V}\left(\int_0^{S_1} r_1^2 (z_1(S_1)-h_0-z_1)\sin(\phi_1)ds_1 
+ \int_0^{S_2} r_2^2 (z_2-h_0-z_2(S_2))\sin(\phi_2)ds_2\right), \nonumber\\
\end{eqnarray}
where $V=V_1+V_2$ is the total volume of the drop and all other variables have been introduced in Section\,\ref{stat}. After some simplification we obtain
\begin{eqnarray}
\label{azc2}
z_c=\frac{V_1z_1(S_1)-\int_0^{S_1}\pi r_1^2z_1\sin(\phi_1)ds_1-V_2z_2(S_2)+\int_0^{S_2}\pi r_2^2z_2\sin(\phi_2)ds_2}{V_1+V_2}-h_0.
\end{eqnarray}

The excess potential energy $U_e$ from Eq.\,(\ref{ue}) of the fluid bath is given by the potential energy of the shaded part in Fig.\ref{F3}, filled with fluid $2$ with the negative density $-\rho_2$. $U_e$ can be split into the energy of the pendant part of the drop and the energy of the remaining upper part of the shaded region in Fig.\ref{F3}
\begin{eqnarray}
\label{aU}
U_e=-g\rho_2\left[ \int_0^{S_2} \pi r_2^2 (z_2-h_0-z_2(S_2))\sin(\phi_2)\,ds_2 +\int_R^\infty \pi r^2 (h(r)-h_0)h'(r)\,dr \right], \nonumber\\
\end{eqnarray}
where $R$ is the contact radius.
Finally, the stability function $f(g)$ is
\begin{eqnarray}
\label{af}
f(g)=z_c+\frac{U_e}{gV\rho_1}.
\end{eqnarray}


\end{document}